\documentclass[aps, pra, preprint, showpacs, superscriptaddress]{revtex4}
\usepackage{amsmath, amssymb, bm, graphicx, CJK, color}
\usepackage{multirow}
\usepackage{dcolumn}
\usepackage[normalem]{ulem}
\begin{document}
\begin{CJK*}{UTF8}{}

\title{Interplay between relativistic energy corrections and resonant excitations in x-ray multiphoton ionization dynamics of Xe atoms}

\author{Koudai Toyota \CJKfamily{min}(豊田広大)}
\email{koudai.toyota@cfel.de}
\affiliation{Center for Free-Electron Laser Science, DESY, 22607 Hamburg, Germany}

\author{Sang-Kil Son \CJKfamily{mj}(손상길)}
\email{sangkil.son@cfel.de}
\affiliation{Center for Free-Electron Laser Science, DESY, 22607 Hamburg, Germany}

\author{Robin Santra}
\email{robin.santra@cfel.de}
\affiliation{Center for Free-Electron Laser Science, DESY, 22607 Hamburg, Germany}
\affiliation{Department of Physics, University of Hamburg, 20355 Hamburg, Germany}

\date{\today}

\begin{abstract}
In this paper, we theoretically study x-ray multiphoton ionization dynamics of heavy atoms taking into account relativistic and resonance effects.
When an atom is exposed to an intense x-ray pulse generated by an x-ray free-electron laser (XFEL), it is ionized to a highly charged ion via a sequence of single-photon ionization and accompanying relaxation processes, and its final charge state is limited by the last ionic state that can be ionized by a single-photon ionization.
If x-ray multiphoton ionization involves deep inner-shell electrons in heavy atoms, energy shifts by relativistic effects play an important role in ionization dynamics, as pointed out in [Phys.\ Rev.\ Lett.\ \textbf{110}, 173005 (2013)].
On the other hand, if the x-ray beam has a broad energy bandwidth, the high-intensity x-ray pulse can drive resonant photo-excitations for a broad range of ionic states and ionize even beyond the direct one-photon ionization limit, as first proposed in [Nature\ Photon.\ \textbf{6}, 858 (2012)].
To investigate both relativistic and resonance effects, we extend the \textsc{xatom} toolkit to incorporate relativistic energy corrections and resonant excitations in x-ray multiphoton ionization dynamics calculations.
Charge-state distributions are calculated for Xe atoms interacting with intense XFEL pulses at a photon energy of 1.5~keV and 5.5~keV, respectively.
For both photon energies, we demonstrate that the role of resonant excitations in ionization dynamics is altered due to significant shifts of orbital energy levels by relativistic effects.
Therefore it is necessary to take into account both effects to accurately simulate multiphoton multiple ionization dynamics at high x-ray intensity.

\end{abstract}


\maketitle
\end{CJK*}

\section{Introduction}

The interaction between x-ray photon and matter is characterized by photoionization and accompanying relaxation processes.
X-ray photoionization predominantly creates a hole in an inner shell of atoms, which subsequently relaxes via Auger decay or fluorescence.
A series of decay processes, a so-called decay cascade, can occur if the hole is formed in a deep inner shell~\cite{CH1966,M1986,MT1987,KD1994,EL2005}.
The interaction with x rays becomes rather complex when the sample is exposed to the unprecedentedly high fluence generated by x-ray free-electron lasers (XFELs)~\cite{PM2016,BB2016,YT2013,FK2013}.
Beyond the one-photon saturation fluence, which is the inverse of the single-photon ionization cross section~\cite{SY2016}, a single atom can absorb more than one photon sequentially after or during decay cascades, and then it becomes a highly charged ion.
In this sequential multiphoton ionization model, the final charge is determined by the last ionic state that can be ionized by one-photon ionization.
This is called ``direct one-photon ionization limit.''

The straightforward sequential multiphoton ionization model has been solved with a rate-equation approach~\cite{RS2007,YK2010} and verified with a series of XFEL experiments on gas-phased atoms: light atoms such as Ne~\cite{YK2010,DR2011} and Ar~\cite{MF2013}, and heavy atoms such as Kr~\cite{RR2013} and Xe~\cite{RS2012,FS2013,MF2013}.
For example, the calculated charge-state distribution (CSD) of Xe at 2~keV showed an excellent agreement with experimental data~\cite{RS2012}.
For Xe at 1.5~keV, however, the highest charge of the experimental CSD exceeded by far the theoretical prediction of the direct one-photon ionization limit~\cite{RS2012}.
To explain this discrepancy, it has been proposed that multiple resonant excitations followed by Auger-like decays, combined with the broad energy bandwidth of SASE XFEL pulses \cite{PM2016}, can drive further ionization beyond the direct one-photon ionization limit.
This is called resonance-enabled or resonance-enhanced x-ray multiple ionization (REXMI)~\cite{RS2012,RR2013} and this mechanism has been implemented in Refs.~\cite{HB2014,HK2015}.
Also the importance of single resonant excitation was found in the study of the Ne atom experimentally~\cite{YK2010} and theoretically~\cite{XG2012}.

When an x-ray photon with higher energy interacts with a heavy atom such as xenon, a deeper inner-shell electron is ionized.
It induces not only more complicated ionization dynamics, but also it is expected that deep inner-shell energy levels are shifted and split due to relativistic effects.
A recent study of Xe at 5.5~keV hinted at this, showing fairly good agreement between theory and experiment, but the theoretical prediction underestimated the yields of highly charged ions due to lack of relativistic effects~\cite{FS2013}.
To the best of our knowledge, a theoretical demonstration of the impact of relativistic effects on x-ray multiphoton ionization dynamics has not been reported yet.

Here we present how relativistic energy corrections and resonant excitations influence x-ray multiphoton ionization dynamics.
For this purpose, we extend the \textsc{xatom} toolkit~\cite{xatom,SY2011,JS2016}.
Our implementation allows us to turn on and off the relativistic effect and the resonance effect, separately.
This allows us to explore the interplay between both classes of effects on x-ray multiphoton ionization dynamics.

This paper is organized as follows. 
In Sec.~\ref{sec:theory}, we introduce our basic equations and notations, and then we formulate relativistic energy corrections, cross sections, and rates. 
We also discuss how to simulate x-ray multiphoton ionization dynamics by using a Monte-Carlo approach~\cite{SS2012}, emphasizing the numerical challenge to handle an extremely large configurational space when resonant excitations are included.
In Sec.~\ref{sec:results}, we present calculations of CSDs for Xe at 1.5~keV and 5.5~keV, which correspond to XFEL experiments conducted at LCLS~\cite{RS2012} and SACLA~\cite{FS2013}, respectively, and discuss both relativistic and resonance effects on x-ray multiphoton ionization dynamics.
In Sec.~\ref{sec:conclusion}, we conclude and give future perspectives.

\section{Theory and numerical details}\label{sec:theory}

In this section, we start with the non-relativistic Hartree-Fock-Slater (HFS) equation and provide relativistic energy corrections to the orbital energy levels.
These are used to calculate photoionization cross sections, Auger rates, fluorescence rates, and resonant photoexcitation cross sections including relativistic effects.
This relativistic framework is implemented as an extension of the \textsc{xatom} toolkit~\cite{xatom,SY2011,JS2016}.
The basic framework can be found in Ref.~\cite{R2009}.
All formulas below can be derived with standard angular-momentum algebra~\cite{VM1988}.
At the end of this section, we will discuss numerical challenges to simulate x-ray multiphoton ionization dynamics when taking into consideration relativistic and resonance effects.

\subsection{Non-relativistic Hartree-Fock-Slater equation}\label{sec:theory-HFS}

We start from a non-relativistic treatment based on the Hartree-Fock-Slater (HFS) method.
The effective one-electron Schr\"odinger equation for an atom is 
\begin{equation}
\label{eq:HFS}
\left[-\frac{1}{2}\nabla^2+V(\mathbf{r})\right]\psi(\mathbf{r})
=\varepsilon \psi(\mathbf{r}).
\end{equation}
Atomic units are used unless specified otherwise.
The potential $V(\mathbf{r})$ is written as
\begin{equation}\label{eq:potential}
V(\mathbf{r})=-\frac{Z}{r}
+\int \! d^3 r' \, \frac{\rho(\mathbf{r}')}{\left|\mathbf{r}-\mathbf{r}'\right|}
+V_\text{x}(\mathbf{r}),
\end{equation}
where $Z$ is the nuclear charge of the atom, and the exchange potential $V_\text{x}(\mathbf{r})$ is approximated by the Slater exchange potential~\cite{S1951},
\begin{equation}
V_\text{x}(\mathbf{r})=-\frac{3}{2}\left[\frac{3}{\pi}\rho(\mathbf{r})\right]^{1/3}.
\end{equation}
The electronic density $\rho(\mathbf{r})$ is given by
\begin{equation}
\rho(\mathbf{r})=\sum_{i}^{N_\text{elec}} \psi_i^\dagger(\mathbf{r}) \psi_i(\mathbf{r}),
\end{equation}
where $\psi_i(\mathbf{r})$ represents the spin-orbital of the $i$th electron,
and the summation runs over the number of electrons $N_\text{elec}$. 
We employ a spherically symmetric electronic density, i.e., $\rho({\mathbf r}) \to \rho(r)$,  
then the potential becomes also spherically symmetric: $V({\mathbf r}) \to V(r)$.
In addition, we use the Latter tail correction~\cite{L1955} to obtain the proper long-range potential 
for both occupied and unoccupied orbitals.
Hence, the spherically symmetric potential is given by
\begin{equation}
V(r) = 
\begin{cases}
-\frac{Z}{r} + \int \! d^3 r' \, \frac{\rho(r')}{\left|\mathbf{r}-\mathbf{r}'\right|} - \frac{3}{2}\left[\frac{3}{\pi}\rho(r)\right]^{1/3} & \text{for }r < r_c,
\\
-\frac{Z - N_\text{elec} + 1}{r} & \text{for }r \geq r_c,
\end{cases}
\end{equation}
where $r_c$ is determined such that $V(r)$ is continuous at $r = r_c$.

Let us consider a spin-orbital $\psi_i(\mathbf{r})$ given in $nlm_lm_s$ representation, 
where $n$, $l$, $m_l$, and $m_s$ are the principal quantum number, the orbital angular momentum
quantum number, the associated projection quantum number, and the spin magnetic quantum number, respectively.
It is then written as
\begin{equation}\label{eq:wf_nlmlms}
\psi_{nlm_lm_s}(\mathbf{r})=\frac{u_{nl}(r)}{r}Y_{l m_l}(\Omega)\chi_{m_s},
\end{equation}
where $Y_{lm_l}(\Omega)$ is a spherical harmonic and $\chi_{m_s}$ is a Pauli two-component spinor.
The function $u_{nl}(r)$ satisfies the radial part of the HFS equation of Eq.~(\ref{eq:HFS}),
\begin{equation}
\label{eq:radial-HFS}
\left[-\frac{1}{2}\frac{d^2}{dr^2}+\frac{l(l+1)}{2r^2}+V(r)\right]u_{nl}(r)=\varepsilon_{nl}u_{nl}(r).
\end{equation}

To solve the HFS eigenvalue problem of Eq.~(\ref{eq:radial-HFS}), 
we employ the generalized pseudospectral (GPS) method based on
the Legendre polynomials~\cite{TC1999}. 
Let $r_\text{max}$ be the maximum radius in the numerical calculation.
The spatial coordinate $r \in [0,r_\text{max}]$ is then mapped onto the finite range
$x \in [-1,1]$ by the following relation,
\begin{equation}
r=L\frac{1+x}{1-x+ 2L / r_\text{max} },
\label{eq:GPS}
\end{equation}
where $L$ is a mapping parameter to tune the distribution of the grid points. 
A small value of $L$ gives us dense grid points near the origin, and a large value of $L$ gives a more uniform distribution of grid points in the interval $[0,r_\text{max}]$.
The solutions are required to satisfy the boundary conditions $u_{nl}(0) = u_{nl}(r_\text{max}) = 0$.
The wave functions of initially occupied orbitals are localized near the origin.
On the other hand, multiple resonant excitations can excite several electrons to high-$n$ Rydberg states that tend to have large amplitudes away from the origin. 
An optimal $L$ needs to be chosen to accurately describe both of them.

\subsection{Relativistic energy corrections}\label{sec:theory-ene-corr}
We treat relativistic effects within first-order degenerate perturbation theory~\cite{HS1963}.
The effective one-body relativistic Hamiltonian for two-component spinors is given by~\cite{BD1964}
\begin{equation}
\hat{H} = \hat{H}_0 + \hat{H}_\text{mass} + \hat{H}_\text{dar} + \hat{H}_\text{so},
\end{equation}
where $\hat{H}_0$ is the non-relativistic Hamiltonian given in Eq.~(\ref{eq:HFS}) as the unperturbed term,
\begin{equation}
\hat{H}_0 = -\frac{1}{2} \nabla^2 + V(r).
\end{equation}
The rest ($\hat{H}' = \hat{H}_\text{mass} + \hat{H}_\text{dar} + \hat{H}_\text{so}$) are the leading-order relativistic corrections, which we treat as perturbations.
The relativistic energy corrections are calculated with these perturbative terms and the non-relativistic spin-orbitals of Eq.~(\ref{eq:wf_nlmlms}).
\begin{subequations}
\label{eq:ene-corr-def}
The first is the mass term, 
\begin{equation}
\label{eq:mass-term}
\hat{H}_\text{mass}=-\frac{\alpha^2}{8}{\hat{p}^4},
\end{equation}
where $\hat{p}$ is the canonical momentum operator and $\alpha$ is the fine structure constant ($\alpha = 1/c$, where $c$ is the speed of light). 
The mass term represents the relativistic mass correction for an electron orbiting at a speed close to $c$.
The second is the Darwin term,
\begin{equation}
\label{eq:dar-term}
\hat{H}_\text{dar}=-\frac{\alpha^2}{4}\frac{dV}{dr}\frac{d}{d r},
\end{equation}
which may be interpreted in terms of Zitterbewegung~\cite{FW1950}.
The last is the spin-orbit coupling term,
\begin{equation}
\label{eq:so-term}
\hat{H}_\text{so}= \frac{\alpha^2}{2}\frac{1}{r}\frac{dV}{dr}{\hat l}\cdot{\hat s}. 
\end{equation}
\end{subequations}

Let the operator $\hat j$ be the sum of the orbital angular momentum operator $\hat l$ and the spin angular momentum operator $\hat s$, then $\hat j=\hat l + \hat s$. 
In the following, we introduce the $nljm$ representation of a spin-orbital,
\begin{equation}
\label{eq:nljm}
\phi_{nljm}(\mathbf{r})=\sum_{m_l m_s}C(lsj;m_lm_sm)\psi_{nlm_lm_s}(\mathbf{r}),
\end{equation}
to calculate relativistic energy corrections and rates.
Here the symbol $C(lsj;m_lm_sm)$ represents a Clebsch-Gordan coefficient $(s=\tfrac{1}{2})$. 
The spin-orbital $\phi_{nljm}(\mathbf{r})$ is a 
simultaneous eigenfunction of  
${\hat l}^2$, ${\hat s}^2$, ${\hat j}^2$, and ${\hat j_z}$. 
Let $\hat O$ be an operator that is independent of angular degrees 
of freedom. Then,
\begin{subequations}
\label{eq:matrix-element}
\begin{equation}
\int \! d^3 r \, \phi^\dagger_{nljm}(\mathbf{r}){\hat O}\phi_{nlj^\prime m^\prime}(\mathbf{r}) =
\langle \hat O \rangle_{nl}\delta_{jj^\prime}\delta_{mm^\prime},
\end{equation}
where
\begin{equation}
\langle \hat O \rangle_{nl}
=\int_0^\infty \! dr \, r^2 \left(\frac{u^*_{nl}(r)}{r}\right){\hat O}\left(\frac{u_{nl}(r)}{r}\right),
\end{equation}
\end{subequations}
and $u_{nl}(r)$ is an eigenfunction of Eq.~(\ref{eq:radial-HFS}).
We evaluate the matrix elements of Eqs.~(\ref{eq:ene-corr-def})
in the $nljm$ representation of Eq.~(\ref{eq:nljm}) with the use of
Eq.~(\ref{eq:radial-HFS}).
Then, the relativistic energy shifts are given by
\begin{subequations}
\label{eq:ene-corr}
\begin{eqnarray}
\Delta \varepsilon^\text{mass}_{nl}
&=&\frac{\alpha^2}{2}\langle \left(\varepsilon_{nl}-V(r)\right)^2 \rangle_{nl}, \\
\Delta \varepsilon^\text{dar}_{nl}
&=&-\frac{\alpha^2}{4}\left \langle \frac{dV}{dr}\frac{d}{dr} \right\rangle_{nl}, \\
\Delta \varepsilon^\text{so}_{nlj}
&=&
\begin{cases}
0 & \text{for }j=\frac{1}{2}, \\
\frac{\alpha^2}{4}l \left \langle \frac{1}{r}\frac{dV}{dr} \right \rangle_{nl} & \text{for }j=l+\frac{1}{2}, \\
-\frac{\alpha^2}{4}(l+1) \left \langle \frac{1}{r}\frac{dV}{dr} \right \rangle_{nl} & \text{for }j=l-\frac{1}{2}.  
\end{cases}
\end{eqnarray}
\end{subequations}
Therefore, an orbital energy level including relativistic energy corrections in the $nljm$ representation is given by
\begin{equation}
\label{eq:rel-ene}
E_{nlj}=\varepsilon_{nl}
+\Delta \varepsilon^{\rm mass}_{nl}
+\Delta \varepsilon^{\rm dar}_{nl}
+\Delta \varepsilon^{\rm so}_{nlj}.
\end{equation} 

Table~\ref{tab:rel-ene} compares the orbital energy levels of neutral Xe including
relativistic energy corrections calculated using Eq.~(\ref{eq:rel-ene}) with those obtained
by Lu \textit{et al.}\ solving the Dirac-Fock-Slater (DFS) equation~\cite{LC1971}.
Our results shown in the table were obtained using $N$=150, $r_\text{max}$=50~a.u., and $L$=10, 
where $N$ is the number of grid points used in the GPS method.
With this parameter set, our calculated values of the mass term in Eq.~(\ref{eq:mass-term}) and the Darwin term in Eq.~(\ref{eq:dar-term}) for the $1s_{1/2}$ orbital deviate $6.3\%$ and $8.0\%$ from those (not shown) in Ref.~\cite{HS1963}, respectively.
Substituting these values into Eq.~(\ref{eq:rel-ene}), the relativistic $1s_{1/2}$ orbital energy agrees with that in Refs.~\cite{HS1963,LC1971,HM1976} to within $0.2\%$. 
Overall, our calculated orbital energies agree with the reference values to within better than 10\%.
The same set of numerical parameters is used in the following subsections for calculating cross sections and rates.

\begin{table}
\caption{\label{tab:rel-ene}Comparison of orbital energy levels of neutral Xe atom (in eV).
``non-rel'' refers to non-relativistic calculations obtained from Eq.~(\ref{eq:radial-HFS}),
and ``rel'' means relativistic calculations using Eq.~(\ref{eq:rel-ene}). 
DFS indicates the Dirac-Fock-Slater results taken from Ref.~\cite{LC1971} and EXP is the experimental data taken from Ref.~\cite{T2009}.} 
\begin{ruledtabular}
\begin{tabular}{crrrrrrr} 
\multicolumn{1}{c}{orbital} &
{non-rel} &
{$E_\text{mass}$} &
{$E_\text{dar}$} &
{$E_\text{so}$} &
{rel} &
{DFS~\cite{LC1971}} & 
{EXP~\cite{T2009}} \\
\hline
$1s_{1/2}$  & 
$-$33102.77 & 
$-$7070.30  &  5565.00 &   0   &  $-$34608.06 & 
$-$34555.26 & $-$34561\phantom{.0}  \\ 
\hline
$2s_{1/2}$  & 
$-$5057.17  & 
$-$972.53  &  587.98  &   0   &  $-$5441.71 & 
$-$5417.15  & $-$5453\phantom{.0}  \\  
\hline 
$2p_{1/2}$ & 
\multirow{2}{*}{$-$4771.10}  & 
\multirow{2}{*}{$-$168.53}   & \multirow{2}{*}{$-$2.46} & $-$199.98  &  $-$5141.09 &  
$-$5104.13 & $-$5107\phantom{.0}    \\
$2p_{3/2}$  &
& & & 99.49 & $-$4842.60 &  
$-$4774.49  & $-$4786\phantom{.0}  \\ 
\hline
$3s_{1/2}$  
& \multirow{1}{*}{$-$1045.69}  
&  $-$205.99 & 117.54 & 0 &  $-$1134.15 
& $-$1122.22 & $-$1148.7 \\
\hline
$3p_{1/2}$  & 
\multirow{2}{*}{$-$922.75} & 
\multirow{2}{*}{$-$42.87} & \multirow{2}{*}{$-$0.46}     & $-$37.84   &  $-$1003.94  & 
$-$989.73 & $-$1002.1  \\
$3p_{3/2}$  &   
& & & 18.92 & $-$947.18 & 
$-$926.51   & $-$940.6  \\ 
\hline
$3d_{3/2}$  & 
\multirow{2}{*}{$-$692.12}  & 
\multirow{2}{*}{$-$11.21}   & \multirow{2}{*}{$-$0.28}   & $-$8.08    & $-$711.71  & 
$-$690.88 	& $-$689.0   \\
$3d_{5/2}$  & 
& & & 5.39  & $-$698.24  & 
$-$677.35   & $-$676.4  \\
\hline
$4s_{1/2}$  & 
$-$192.78   & 
$-$46.02    &  25.82 & 0 & $-$212.99 & 
$-$208.50   &  $-$213.2   \\
\hline
$4p_{1/2}$  & 
\multirow{2}{*}{$-$148.44} & 
\multirow{2}{*}{$-$9.32}   & \multirow{2}{*}{$-$0.09}    & $-$7.65     & $-$165.52 & 
$-$160.81 & $-$146.7\\
$4p_{3/2}$  &
& & & 3.82  & $-$154.03 & 
$-$148.00   & $-$145.5   \\
\hline
$4d_{3/2}$  & 
\multirow{2}{*}{$-$71.42}   & 
\multirow{2}{*}{$-$2.17}    & \multirow{2}{*}{$-$0.04}  & $-$1.28  & $-$74.93& 
$-$69.85 	& $-$69.5  \\
$4d_{5/2}$  &
& & &  0.85      & $-$72.79 & 
$-$67.77    & $-$67.5   \\
\hline
$5s_{1/2}$  & 
\multirow{1}{*}{$-$21.78} & 
$-$6.84 & 3.82 &  0  & $-$24.80 & 
$-$23.65 	& $-$23.3    \\
\hline
$5p_{1/2}$  & 
\multirow{2}{*}{$-$11.39}  & 
\multirow{2}{*}{$-$1.05}   & \multirow{2}{*}{$-$0.01}  & $-$0.84 & $-$13.30 & 
$-$12.39 	& $-$13.4 \\
$5p_{3/2}$  &
& & & 0.42      & $-$12.03 & 
$-$10.97 	& $-$12.1 \\
\end{tabular}
\end{ruledtabular}
\end{table}

\subsection{Photoionization cross section}\label{sec:theory-pcs}
The photoionization cross section of a subshell $i$ at incident photon energy $\omega_{\rm in}$ is given by
\begin{equation}
\label{eq:pcs}
\sigma_p(i,\omega_{\rm in})=
\frac{4}{3}\pi^2\alpha\omega_{\rm in}N_i\sum_{l_a = |l_i \pm 1|}
\frac{l_>}{2l_i+1}|\langle u_{E_al_a}|r|u_{n_il_i} \rangle|^2,
\end{equation}
where $N_i$ is the number of electrons in the subshell $i$,
and $l_>$ is the greater of $l_i$ and $l_a$. The energy $E_a ( > 0 )$ of a 
final state $a$ is given by $E_a=E_{n_il_ij_i}+\omega$, where
the quantity $E_{n_il_ij_i}$ is an orbital energy level including
relativistic energy corrections, Eq.~(\ref{eq:rel-ene}). 
Here, $u_{E_al_a}$ is the continuum state with the positive energy $E_a$, 
which is computed using the fourth-order Runge-Kutta method on a uniform grid~\cite{C1962,MC1968}. 
The selection rules are 
\begin{subequations}
\label{eq:dip-selection}
\begin{eqnarray}
j_a &=& j_i,~j_i\pm 1, \\
l_a &=& l_i \pm 1.
\end{eqnarray}
\end{subequations}

Table~\ref{tab:pcs} compares our results with the values calculated by using DFS~\cite{S1973}.
The deviations between them are found to be larger for deep inner shells. 
Since relativistic energy corrections lower orbital energy levels, relativistic photoionization 
cross sections summed over different $j$'s become bigger than those in the non-relativistic case.
Note that Table~\ref{tab:pcs} shows photoionization cross sections of the ground configuration of neutral Xe atom only.
For most multiple-hole states arising as a consequence of strong x-ray exposure, as discussed in Sec.~\ref{sec:rate_equation}, one cannot simply consult the literature.

\begin{table}
\caption{\label{tab:pcs}Comparison of photoionization cross sections of neutral 
Xe atom at 5455 keV (in kbarns).  ``non-rel'' refers to non-relativistic
calculations using Eq.~(6) in Ref.~\cite{SY2011}, and ``rel'' is relativistic calculations
using Eq.~(\ref{eq:pcs}). 
DFS results are taken from Ref.~\cite{S1973}. }
\begin{ruledtabular}
\begin{tabular}{crrr}
orbital         & non-rel                     & rel           & DFS~\cite{S1973} \\
\hline
$2s_{1/2}$       & 23.96                       & 28.05         & 23.37 \\
\hline
$2p_{1/2}$       & \multirow{2}{*}{110.56}     & 47.69         & 40.96 \\
$2p_{3/2}$       &                             & 77.39         & 73.73 \\
\hline 
$3s_{1/2}$       & 5.49                        & 5.76          & 5.54 \\
\hline
$3p_{1/2}$       & \multirow{2}{*}{15.14}      & 5.31          & 5.75  \\
$3p_{3/2}$       &                             & 10.25         & 10.34 \\
\hline
$3d_{3/2}$       & \multirow{2}{*}{8.28}       & 3.36          & 3.56 \\
$3d_{5/2}$       &                             & 4.99          & 4.95 \\  
\hline
$4s_{1/2}$       & 1.24                        & 1.25          & 1.26 \\
\hline
$4p_{1/2}$       & \multirow{2}{*}{2.96}       & 0.99          & 1.13 \\
$4p_{3/2}$       &                             & 1.98          & 2.02 \\
\hline
$4d_{3/2}$       & \multirow{2}{*}{1.34}       & 0.54          & 0.57 \\
$4d_{5/2}$       &                             & 0.80          & 0.79 \\
\hline
$5s_{1/2}$       & 0.18                        & 0.18          & 0.19 \\
\hline
$5p_{1/2}$       & \multirow{2}{*}{0.32}       & 0.10          & 0.12 \\
$5p_{3/2}$       &                             & 0.21          & 0.21 \\
\end{tabular}
\end{ruledtabular}
\end{table}

\subsection{Auger rate}\label{sec:theory-Auger}

The Auger rate at which an initial hole in a subshell $i$ 
is refilled by an electron from a subshell $q$ or $q^\prime$
accompanied by the emission of an Auger electron from the subshell 
$q$ or $q^\prime$ is given by~\cite{A1963}
\begin{equation}
\label{eq:auger}
\Gamma_{i,qq^\prime}=2\pi N_i^{\rm H}N_{q q^\prime}
\sum_{l_a=0}^{l_i+l_q+l_{q^\prime}}
\sum_{j_a=|l_a-\frac{1}{2}|}^{l_a+\frac{1}{2}}
\sum_{J=|j_q-j_{q^\prime}|}^{j_q+j_{q^\prime}}
(2j_a+1)(2J+1)|M_J(aiqq^\prime)|^2.
\end{equation}
The kinetic energy of the Auger electron is given by 
$E_a=-E_{n_il_ij_i}+E_{n_ql_qj_q}+E_{n_{q^\prime}l_{q^\prime}j_{q^\prime}}$.
The constant $N_i^{\rm H}$ is the number of initial holes in the subshell $i$.
Let $N_{q}$ be the number of electrons in the subshell $q$, then
the quantity $N_{qq^\prime}$ is defined by
\begin{equation}
N_{qq^\prime}=
\begin{cases}
N_{q}N_{q^\prime} & \text{(inequivalent~electrons)}, \\
\frac{2j_q+1}{2j_q}N_{q}(N_q-1) & \text{(equivalent~electrons)}.
\end{cases}
\end{equation}
The function $M_J(aiqq^\prime)$ is defined by
\begin{subequations}
\begin{equation}
M_J(aiqq^\prime)=\tau \sum_{k=0}^{l_i+l_q+l_{q^\prime}}
\left[
R_k(aiqq^\prime)A_{kJ}(aiqq^\prime)+(-1)^{-J}R_k(aiq^\prime q)A_{kJ}(aiq^\prime q)
\right],
\end{equation}
where the coefficient $\tau$ is given by
\begin{equation}
\tau=
\begin{cases}
1 & \text{(inequivalent~electrons)}, \\
\frac{1}{\sqrt{2}} & \text{(equivalent electrons)}.
\end{cases}
\end{equation}
The functions $R_k(aiqq^\prime)$ is defined by
\begin{equation}
R_k(aiqq^\prime)=\int_0^\infty \! dr \, \int_0^\infty \! dr^\prime \, 
u_{E_al_a}(r)u_{n_il_i}(r^\prime)
\frac{r_<^k}{r_>^{k+1}}u_{n_ql_q}(r)u_{n_{q^\prime} l_{q^\prime}}(r^\prime),
\end{equation}
where $r_>~(r_<)$ is the greater (lesser) of $r$ and $r^\prime$,  
and the function $A_{kJ}(aiqq^\prime)$ is defined by 
\begin{equation}
\label{eq:akj}
A_{kJ}(aiqq^\prime)
=
\left\{
\begin{array}{ccc}
j_a & j_q & k \\
j_{q^\prime} & j_i & J 
\end{array}
\right\}
\left\{
\begin{array}{ccc}
l_q & s & j_q \\
j_a & k & l_a 
\end{array}
\right\}
\left\{
\begin{array}{ccc}
l_{q^\prime} & s & j_{q^\prime} \\
j_i & k & l_i 
\end{array}
\right\}
\langle l_a||C_k||l_q \rangle
\langle l_i||C_k||l_{q^\prime} \rangle,
\end{equation}
where the braces are $6j$ symbols, and
\begin{equation}
\langle l^\prime||C_k||l \rangle=\sqrt{2k+1}C(lkl^\prime;000).
\end{equation}
\end{subequations}

\subsection{Fluorescence rate}\label{sec:theory-fluo}
The fluorescence rate at which an initial hole in a subshell $q$ 
is refilled by an electron from a subshell $q^\prime$ 
accompanied by the emission of a photon 
is given by 
\begin{equation}
\label{eq:fluo}
\Gamma_{qq^\prime}=\frac{4}{3}
\alpha^3 (E_{n_{q^\prime}l_{q^\prime}j_{q^\prime}}-E_{n_ql_qj_q})^3
l_>N_{q^\prime} N_{q}^{\rm H}
\left\{
\begin{array}{ccc}
l_{q^\prime} & s& j_{q^\prime} \\
j_q & 1 & l_q
\end{array}
\right\}^2
\left|\langle u_{n_ql_q}|r|u_{n_{q^\prime}l_{q^\prime}} \rangle \right|^2,
\end{equation}
where $N_{q^\prime}$ and $N_{q}^{\rm H}$ represent the number of
electrons and holes in the subshells $q^\prime$ and $q$, respectively.
Here it is assumed that the emitted photon is not polarized.
The selection rules are
\begin{subequations}
\begin{eqnarray}
j_q&=&j_{q^\prime},~j_{q^\prime}\pm 1, \\
l_q&=& l_{q^\prime}\pm 1.
\end{eqnarray}
\end{subequations}

\begin{table}
\caption{\label{tab:aug-fluo}Comparison of fluorescence ($L$-$X$ or $M$-$X$), Auger ($L$-$XY$ or $M$-$XY$), and Coster-Kronig ($L$-$LX$ or $M$-$MX$) rates for $M$- and $L$-shell vacancies of Xe (in a.u.).
``non-rel'' refers to the non-relativistic calculation, and ``rel'' is obtained from Eq.~(\ref{eq:auger}) or (\ref{eq:fluo}).
DF refers to the multiconfiguration Dirac-Fock calculations~\cite{SM2015}.
Note that the Coster-Kronig channels of $L_2$-$L_3X$ and $M_2$-$M_3X$ are energetically forbidden in the non-relativistic case.}
\begin{ruledtabular}
\begin{tabular}{lccc}
group           & non-rel				& rel 					& DF~\cite{SM2015} \\
\hline
$L_1$-$X$		& $6.33\times10^{-3}$	& $8.03\times10^{-3}$	& $6.33\times10^{-3}$\\
$L_1$-$XY$      & $6.07\times10^{-2}$	& $5.63\times10^{-2}$	& $6.50\times10^{-2}$\\
$L_1$-$L_{23}X$	& $8.19\times10^{-2}$	& $6.76\times10^{-2}$ 	& $5.83\times10^{-2}$\\
\hline
$L_2$-$X$		& $1.04\times10^{-2}$	& $1.32\times10^{-2}$	& $5.35\times10^{-3}$\\
$L_2$-$XY$      & $9.38\times10^{-2}$	& $8.70\times10^{-2}$	& $4.86\times10^{-2}$\\
$L_2$-$L_3X$	& forbidden             & $2.01\times10^{-2}$ 	& $6.82\times10^{-3}$\\
$L_3$-$X$		& (=$L_2$-$X$)			& $1.08\times10^{-2}$ 	& $1.01\times10^{-2}$\\
$L_3$-$XY$      & (=$L_2$-$XY$)			& $9.28\times10^{-2}$	& $1.10\times10^{-1}$\\
\hline
$M_1$-$X$		& $1.72\times10^{-4}$	& $2.27\times10^{-4}$	& $1.72\times10^{-4}$\\
$M_1$-$XY$      & $1.85\times10^{-2}$	& $1.73\times10^{-2}$	& $1.91\times10^{-2}$\\
$M_1$-$M_{23}X$	& $4.74\times10^{-1}$	& $3.46\times10^{-1}$ 	& $3.72\times10^{-1}$\\
$M_1$-$M_{45}X$	& $8.97\times10^{-2}$	& $8.06\times10^{-2}$ 	& $8.38\times10^{-2}$\\
\hline
$M_2$-$X$		& $1.61\times10^{-4}$	& $2.19\times10^{-4}$	& $1.75\times10^{-4}$\\
$M_2$-$XY$      & $2.09\times10^{-2}$	& $1.97\times10^{-2}$	& $2.10\times10^{-2}$\\
$M_2$-$M_3X$	& forbidden             & $5.39\times10^{-3}$ 	& $6.20\times10^{-4}$\\
$M_2$-$M_{45}X$	& $2.05\times10^{-1}$	& $1.54\times10^{-1}$ 	& $1.64\times10^{-1}$\\
$M_3$-$X$		& (=$M_2$-$X$)			& $1.75\times10^{-4}$ 	& $1.73\times10^{-4}$\\
$M_3$-$XY$      & (=$M_2$-$XY$)			& $2.06\times10^{-2}$ 	& $2.22\times10^{-2}$\\
$M_3$-$M_{45}X$ & (=$M_2$-$M_{45}X$)	& $1.88\times10^{-1}$ 	& $1.76\times10^{-1}$\\  
\hline
$M_4$-$X$		& $1.02\times10^{-5}$	& $1.05\times10^{-5}$	& $1.21\times10^{-5}$\\
$M_4$-$XY$      & $2.25\times10^{-2}$	& $2.23\times10^{-2}$ 	& $2.46\times10^{-2}$\\
$M_5$-$X$		& (=$M_4$-$X$)			& $1.03\times10^{-5}$ 	& $1.05\times10^{-5}$\\
$M_5$-$XY$      & (=$M_4$-$XY$)			& $2.26\times10^{-2}$ 	& $2.16\times10^{-2}$\\
\end{tabular}
\end{ruledtabular}
\end{table}

Table~\ref{tab:aug-fluo} compares our numerical results for Auger (Coster-Kronig) and fluorescence rates with the recent data obtained with the multiconfiguration Dirac-Fock method~\cite{SM2015}.
It is important to realize that the Coster-Kronig channels of $L_2$-$L_3X$ and $M_2$-$M_3X$ 
are completely missing in the non-relativistic case, because they are energetically forbidden without the spin-orbit energy splitting.

\subsection{Resonant photoexcitation cross section}\label{sec:theory-res}

We consider the cross section of a resonant excitation for a 
bound-to-bound transition from an initial to a final orbital, $i \to f$,
\begin{subequations}
\begin{equation}
\sigma_R(i \to f, \omega)=
\frac{4}{3}\pi^2\alpha \omega
l_>N_{i}N_{f}^H
\left\{
\begin{array}{ccc}
l_i & s & j_i \\
j_f & 1 & l_f  
\end{array}
\right\}^2
|\langle u_{n_f l_f}|r|u_{n_il_i} \rangle|^2\delta(\omega-\Delta E_{fi}),
\end{equation}
where the delta function represents the energy conservation law.
The quantity $\Delta E_{fi}$ represents the transition energy given by
\begin{equation}
\Delta E_{fi}=E_{n_fl_fj_f}-E_{n_il_ij_i}.
\end{equation}
\end{subequations}

Assuming that the photon energy spectrum is given by a Gaussian function,
\begin{equation}
\label{eq:fe}
f(\omega;\omega_\text{in})=\frac{1}{\Delta \omega_\text{in}} \sqrt{\frac{4 \ln 2}{\pi}} e^{- 4 \ln 2 \left( \frac{ \omega-\omega_\text{in} }{ \Delta \omega_\text{in} } \right)^2 },
\end{equation} 
where $\Delta \omega_\text{in}$ is the full-width-at-half-maximum (FWHM) of 
the photon-energy distribution function.
Convolving the cross section with the spectral distribution profile of Eq.~(\ref{eq:fe}),
we obtain
\begin{equation}
\label{eq:pcs-res}
\sigma_R(i \to f, \omega_\text{in})
=\frac{4}{3}\pi^2\alpha \Delta E_{fi}
l_>N_{i}N_{f}^H
\left\{
\begin{array}{ccc}
l_i & s & j_i \\
j_f & 1 & l_f  
\end{array}
\right\}^2
|\langle u_{n_f l_f}|r|u_{n_il_i} \rangle|^2f(\Delta E_{fi};\omega_\text{in}).
\end{equation}
Replacing the subscript $a$ with $f$ in Eq.~(\ref{eq:dip-selection}), 
we obtain the selection rules. 

\subsection{Rate equations for ionization dynamics}\label{sec:rate_equation}

We employ a rate-equation approach to simulate x-ray multiphoton ionization 
dynamics~\cite{RS2007,YK2010}.
The eigenfunctions and energies of the HFS equation in Eq.~(\ref{eq:HFS}) 
are used to calculate the cross sections of Eqs.~(\ref{eq:pcs}) and (\ref{eq:pcs-res}) 
and rates of Eqs.~(\ref{eq:auger}) and (\ref{eq:fluo}).
Time-dependent photoionization and photoexcitation rates 
at a given time are calculated by their respective cross sections 
times the photon flux at that time.
All calculated rates are plugged into a set of coupled rate equations,
\begin{equation}
\label{eq:req}
\frac{dP_I}{dt}=\sum_{I^\prime \neq I}^{\rm all~config.}
\left[
\Gamma_{I^\prime \to I}(t) P_{I^\prime}(t)-\Gamma_{I \to I^\prime}(t) P_I(t)
\right],
\end{equation}
where $P_{I}$ is the population of the $I$th electronic configuration and
$\Gamma_{I \to I^\prime}$ is the transition rate from $I$ to $I^\prime$. 

The dimension of the rate-equation system of Eq.~(\ref{eq:req}) becomes enormously large
for heavy atoms. 
For example, let us consider a neutral Xe atom, which has 54 electrons, and construct all possible electronic configurations that may be formed by removing zero, one, or more electrons, from the neutral ground configuration.
All possible configurations of Xe ions, Xe$^{q+}$, in the non-relativistic case are written as
\begin{equation*}
\text{Xe$^{q+}$: } 1s^{n_1} \; 2s^{n_2} 2p^{n_3} \; 3s^{n_4} 3p^{n_5} 3d^{n_6} \; 4s^{n_7} 4p^{n_8} 4d^{n_9} \; 5s^{n_{10}} 5p^{n_{11}},
\end{equation*}
where $n_i$ is chosen from 0 to the maximum occupation number ($n_i^\text{max}$) of the $i$th subshell, i.e., $n_1 = 0, 1, 2$; $n_3 = 0, 1, \cdots, 6$; and so on.
The sum of $\lbrace n_i \rbrace$ gives the total number of electrons: $\sum_i n_i = 54 - q$.
The number of all possible configurations, which is equal to the
number of coupled rate equations that must be solved, 
is given by $N_\text{config} = \prod_i ( n_i^\text{max} + 1 )$.
For the Xe case, it gives $3 \times (3 \times 7) \times (3 \times 7 \times 11) \times (3 \times 7 \times 11) \times (3 \times 7) = 70\;596\;603$.
When relativistic effects are taken into account, $N_\text{config}$ is further increased by about 200 times because of the spin-orbit splittings ($p_{1/2} / p_{3/2}$ and $d_{3/2} / d_{5/2}$), so the number of rate equation becomes $15\;069\;796\;875$.
If resonant bound-to-bound excitations are considered, $N_\text{config}$ explodes (see Table I in 
Ref.~\cite{HK2015}), even without consideration of relativistic effects.

Directly solving such a gigantic number of coupled rate equations 
is thus impractical. Instead, we extend \textsc{xatom} to employ the Monte-Carlo method
to solve Eq.~(\ref{eq:req}) with pre-calculated tables of cross sections and rates, as previously demonstrated in Ref.~\cite{SS2012}.
Furthermore, the electronic structure, cross sections, and rates are calculated on the fly, only when a Monte-Carlo trajectory visits a new electronic configuration~\cite{FS2013}. 
This Monte-Carlo on-the-fly scheme dramatically saves computational effort, enabling us to explore very complicated ionization dynamics of heavy atoms.
A detailed Monte-Carlo description for x-ray multiphoton ionization dynamics is found in Ref.~\cite{SS2012}.
A Monte-Carlo convergence is checked out at every 100 trajectories.
When the absolute differences of charge state populations between current and previous checking points become less than $10^{-4}$ ($10^{-5}$ for the 5.5~keV case in Sec.~\ref{sec:CSD@5.5keV}), the program terminates the Monte-Carlo calculation.

\subsection{Comparison of computational times}

\begin{figure}
\includegraphics[width=0.75\textwidth]{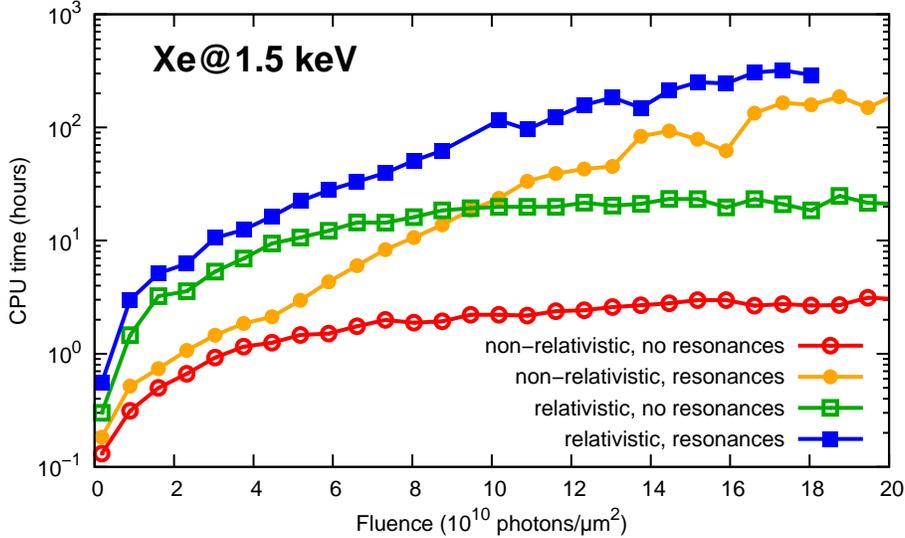} 
\caption{Comparison of CPU times for calculating x-ray multiphoton ionization dynamics of Xe at 1.5~keV
with a bandwidth of 15~eV as a function of fluence. The open circles (red) and open squares (green)   
represent non-relativistic and relativistic calculations excluding resonant excitations. 
The filled circles (orange) and filled squares (blue) represent non-relativistic and relativistic calculations including resonant excitations.}
\label{fig:cputime}
\end{figure}

Now we compare computational times for calculating x-ray multiphoton ionization dynamics, turning on and off relativistic and resonance effects.
Thus, we have four different cases:
(a) \emph{non-rel, nores}: without relativistic effects and without resonant excitations; 
(b) \emph{non-rel, res}: without relativistic effects and with resonant excitations;
(c) \emph{rel, nores}: with relativistic effects and without resonant excitations; and
(d) \emph{rel, res}: with relativistic effects and with resonant excitations.
Figure~\ref{fig:cputime} shows the CPU time for each case as a function of fluence, calculating the charge-state distribution (CSD) of Xe atom at 1.5~keV.
A photon energy bandwidth of 15~eV FWHM is used when resonant excitations are included.
The pulse duration is fixed at 80~fs FWHM.

First, let us examine the cases excluding resonances, (a) and (c).
The relativistic calculations (green, open squares) take longer CPU time by one order of magnitude than the non-relativistic calculations (red, open circles).
In both cases, the computational time saturates as the fluence increases, because the direct one-photon ionization is no longer possible beyond a certain charge state without resonant excitations and all of the Monte-Carlo trajectories are stuck at the direct one-photon ionization limit (Xe$^{26+}$ at 1.5~keV).
On the other hand, for the cases including resonances, (b) and (d), as plotted with filled circles and filled squares, the CPU time keeps increasing as the fluence increases.
This is because resonant excitations open up new ionization channels beyond the direct one-photon ionization limit and Monte-Carlo trajectories take more time to arrive at their final charge state as the fluence increases.
At the highest fluence ($\sim 2\times10^{11}$~photons/$\mu$m$^2$) used in our calculations, the CPU time of the resonant cases is more than one order of magnitude longer than that of the nonresonant cases.
Overall, at the highest fluence, the relativistic calculation with resonant excitations takes $\sim$100 times more CPU time than the non-relativistic calculation without resonant excitations.

It is worthwhile to discuss how to choose computational parameters, especially for calculating the resonant cases.
Resonant excitation involves an electronic transition to a Rydberg state.
The maximum radius $r_\text{max}$, the mapping parameter $L$, and the number of grid points $N$ must be large enough to provide an accurate description of high-$n$ Rydberg states.
Thus, they must be varied and checked to obtain converged CSDs.
At the same time, it is also necessary to restrict $n_\text{max}$ and $l_\text{max}$ 
when calculating Rydberg states.
In principle, the orbital angular momentum quantum number $l$ is given by $0 \leq l \leq n-1$, 
but in practice $l_\text{max}$ can be chosen much smaller than $(n_\text{max}-1)$ for two reasons.
First, single resonant excitation will increase (or decrease) $l$ by $\pm 1$.
Second, REXMI~\cite{RS2012} involves multiply excited states and their autoionization, which could potentially give a high-$l$ state.
For example, a doubly excited state with two $l=3$ excited electrons can decay into an $l=6$ state, 
and another electronic decay from this state can increase $l$ further.
A rule of thumb is to choose $l_\text{max}$ higher than two times the highest $l$ of the initially occupied subshells.
Eventually $l_\text{max}$ as well as $n_\text{max}$ should be chosen as convergence parameters.
We found that $n_\text{max}$=21 and $l_\text{max}$=6 are necessary to get converged results for 
the non-relativistic resonant case in the fluence regime used in Fig.~\ref{fig:cputime}.
Note that those are larger than the values ($n_\text{max}$=10 and $l_\text{max}$=4) used in Refs.~\cite{HB2014,HK2015}.
In Table~{\ref{tab:comp_param}}, we list all computational parameters ($N$, $L$, $r_\text{max}$, $n_\text{max}$, and $l_\text{max}$) that we used for the x-ray multiphoton dynamics calculations presented in the following sections.

\begin{table}
\caption{\label{tab:comp_param}%
Computational parameters used for ionization dynamics calculations}
\begin{ruledtabular}
\begin{tabular}{lrrrrr}
							& $N$	& $L$ (a.u.)	& $r_\text{max}$ (a.u.)	& $n_\text{max}$	& $l_\text{max}$ \\
\hline
Xe at 1.5~keV \\
\hspace{18pt}non-rel, nores	& 150	& 10			& 50					& --				& --	\\
\hspace{18pt}non-rel, res	& 150	& 10			& 100					& 21				& 6		\\
\hspace{18pt}rel, nores		& 150	& 10			& 50					& --				& --	\\
\hspace{18pt}rel, res		& 150	& 10			& 100					& 19				& 7		\\
\hline
Xe at 5.5~keV \\
\hspace{18pt}non-rel, nores	& 150	& 10			& 50					& --				& --	\\
\hspace{18pt}non-rel, res	& 150	& 10			& 100					& 15				& 6		\\
\hspace{18pt}rel, nores		& 150	& 10			& 50					& --				& --	\\
\hspace{18pt}rel, res		& 150	& 10			& 100					& 12				& 7		\\
\end{tabular}
\end{ruledtabular}
\end{table}

\section{Results}\label{sec:results}

\subsection{X-ray multiphoton ionization dynamics of Xe at 5.5 keV}\label{sec:CSD@5.5keV}

\begin{figure}
\includegraphics[width=0.75\textwidth]{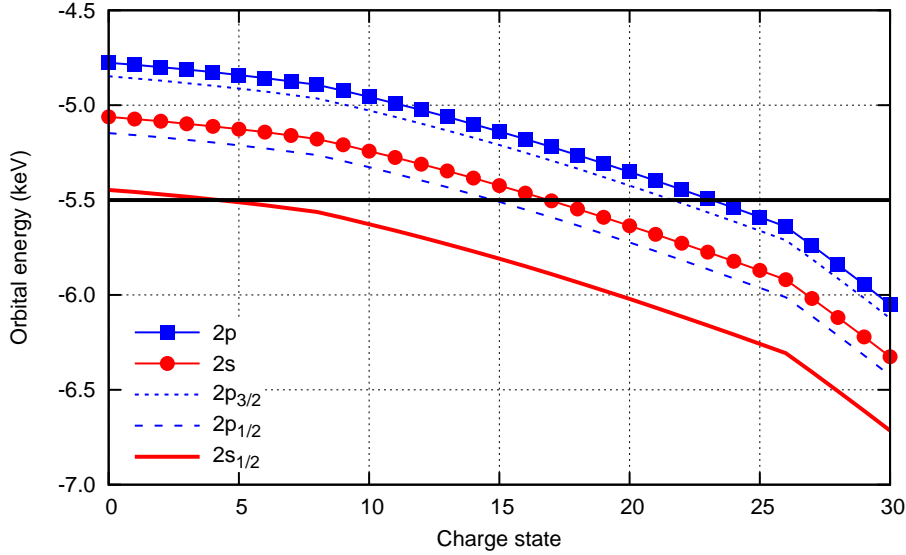} 
\caption{$L$-shell orbital energies of the ground configuration of Xe ions with and without relativistic effects.}
\label{fig:orbital_E@5.5keV}
\end{figure}

We investigate x-ray multiphoton ionization dynamics of Xe at 5.5~keV, where theory predicted lower populations for high charge states in comparison with experimental data~\cite{FS2013}.
It was speculated that this underestimation might be attributed to the lack of relativistic effects and shake-off processes in the theoretical model used.
A 5.5-keV photon can ionize $L$-shell electrons, where the relativistic correction is a few hundred eV and the spin-orbit splitting of $2p$ is about 300~eV (see Table~\ref{tab:rel-ene}).
Figure~\ref{fig:orbital_E@5.5keV} shows the $L$-shell orbital energies of the ground configuration of Xe ions as a function of charge state.
The non-relativistic calculations ($2s$ and $2p$) are plotted with lines and symbols, 
while the relativistic calculations ($2s_{1/2}$, $2p_{1/2}$, and $2p_{3/2}$) are plotted with lines only.
For the whole range of charge states, the $2s_{1/2}$ orbital energies are lower by about 400~eV than the $2s$ orbital energies, and the $2p$ orbital energies are split into the $2p_{1/2}$ and $2p_{3/2}$ by about 300~eV.
Therefore, relativistic energy corrections may affect ionization dynamics in the following ways.
First, the photoionization cross section of neutral Xe at 5.5~keV becomes higher because the ionization potential of $2s_{1/2}$ is closer to the photon energy. 
Second, the spin-orbit splitting allows decay channels that are completely absent in the non-relativistic calculations, for example, the $L_2$--$L_3 X$ and $M_2$--$M_3 X$ channels in Table~\ref{tab:aug-fluo}.
Both effects may enhance the degree of multiple ionization.
Third, the sequence of ionization events in the relativistic case stops earlier than that in the non-relativistic case.
As shown in Fig.~\ref{fig:orbital_E@5.5keV}, $2s$ photoionization at 5.5~keV stops at Xe$^{17+}$ in the non-relativistic case, whereas $2s_{1/2}$ photoionization stops at Xe$^{5+}$ in the relativistic case.
Note that this effect suppresses the yields of higher charge states, unless we consider resonance-driven processes beyond the direct one-photon ionization limit.

\begin{figure}
\includegraphics[width=0.75\textwidth]{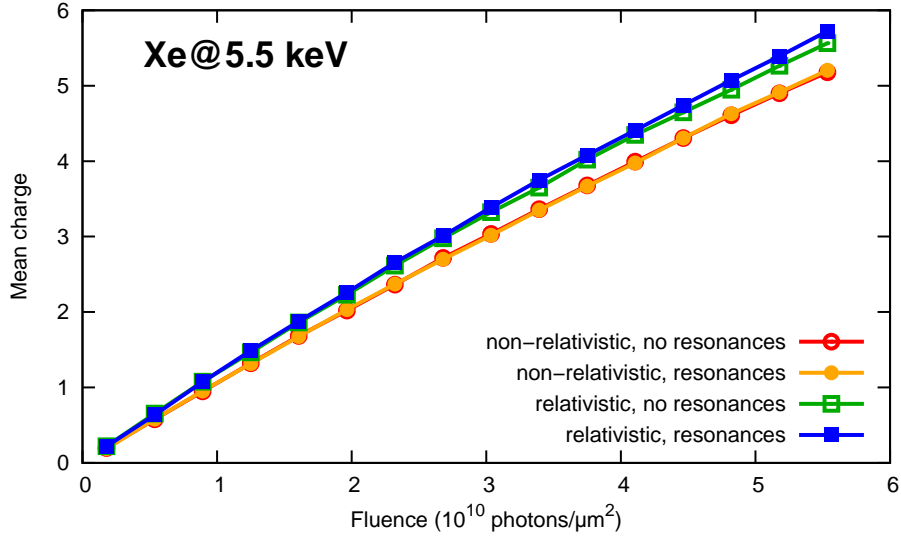} 
\caption{Mean charge of Xe at 5.5~keV as a function of fluence.}
\label{fig:mean_charge@5.5keV}
\end{figure}

We plot the mean charge of Xe at 5.5~keV as a function of fluence for the four different cases (with/without relativistic effects and with/without resonance effect) in Fig.~\ref{fig:mean_charge@5.5keV}.
The pulse duration is 30~fs FWHM.
The spectral bandwidth of the x-ray pulse is assumed to be 1\% (55~eV) for the resonance calculations.
All computational parameters used are listed in Table~\ref{tab:comp_param}.
One can see that the relativistic calculations (green and blue squares) yield higher mean changes than the non-relativistic calculations (red and orange circles).
The difference in the mean charges is due to the difference in photoionization cross sections of neutral Xe at 5.5~keV: 0.166~Mb in the non-relativistic case and 0.186~Mb in the relativistic case.

\begin{figure}
\includegraphics[width=0.75\textwidth]{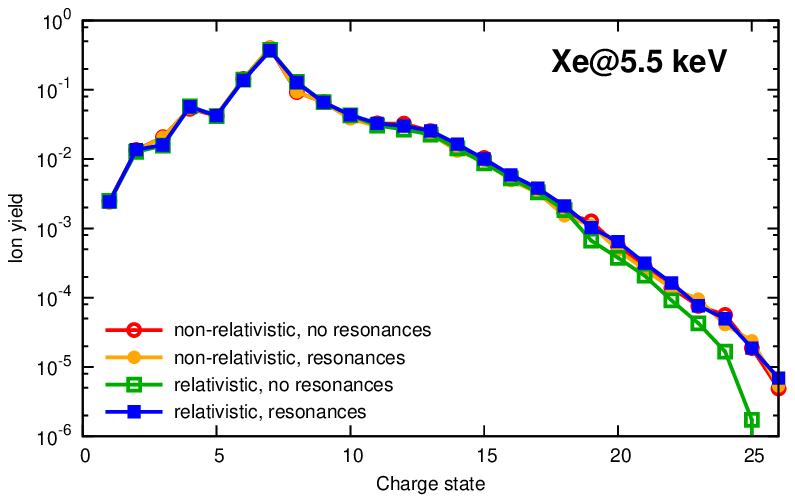} 
\caption{Charge-state distributions of Xe at 5.5~keV, after volume integration with a peak fluence of $5.3\times10^{10}$~photons/$\mu$m$^2$.}
\label{fig:CSD@5.5keV}
\end{figure}

On the other hand, one can see counterintuitive results when the CSDs are examined.
Figure~\ref{fig:CSD@5.5keV} shows Xe CSDs at 5.5~keV for the four different cases.
A peak fluence of $5.3\times10^{10}$~photons/$\mu$m$^2$ is used for volume integration~\cite{YK2010}, assuming a Gaussian beam profile.
All results are similar to each other, except the relativistic calculation without resonances.
Note that the mean charge enhancement of the relativistic case in Fig.~\ref{fig:mean_charge@5.5keV} is not clearly shown in the CSDs because of the logarithmic scale of the ion yields.
It is surprising to see that, without resonances, the relativistic calculation (green, open squares) underestimates the yields of highly charged ions in comparison with the non-relativistic calculation (red, open circles), which is opposite to the expectation from Refs.~\cite{SS2012,FS2013} that those 
for the relativistic calculation would be enhanced.
Since direct one-photon ionization closes at $+5$ in the relativistic case, which is much earlier than $+17$ in the non-relativistic case, the ion yields in higher charge states are suppressed, even though the mean charge is enhanced. 

Next, let us compare the non-relativistic results without resonances (red, open circles) and with resonances (orange, filled circles) in Fig.~\ref{fig:CSD@5.5keV}.
It is worthwhile to note that resonance-driven multiple ionization here is resonance-\emph{enhanced}~\cite{RR2013} (both photoionization and resonant excitation are allowed), not resonance-\emph{enabled}~\cite{RS2012} (which would be the case if only resonant excitation were possible).
In the non-relativistic case, when resonant photoexcitation from $2s$ starts at Xe$^{17+}$, all $2p$ electrons are still available for photoionization.
Since the $2p$ photoionization cross section (0.11~Mb) is much larger than the $2s \rightarrow np$ resonant photoexcitation cross section ($\sim$1~kb), the $2p$ photoionization is the predominant process for ionization.
Therefore, resonant excitation only enhances the ionization yields a bit, in addition to photoionization.
Even when $2p$ photoionization closes at $+23$, many valence electrons are still available for photoionization, and their cross section is similar to the resonant excitation cross section.
Thus, one can see from Fig.~\ref{fig:CSD@5.5keV} that the resonance effect at 5.5~keV at the given peak fluence is almost negligible for the non-relativistic calculations.

The situation is somewhat different when both relativistic and resonance effects are taken into account.
Let us compare the relativistic results without resonances (green, open squares) and with resonances (blue, filled squares) in Fig.~\ref{fig:CSD@5.5keV}.
Since the direct one-photon ionization in the relativistic calculations stops much earlier than that in the non-relativistic calculations, the enhancement effect due to resonant excitation becomes more visible.
One can clearly see that the resonance effect enhances the yields of high charge states in the relativistic case, in contrast to the non-relativistic case.
It turns out that the relativistic, resonant results (blue, filled squares) coincidentally overlap with the non-relativistic results (red and orange circles).

\subsection{X-ray multiphoton ionization dynamics of Xe at 1.5~keV}\label{sec:CSD@1.5keV}

\begin{figure}
\includegraphics[width=0.75\textwidth]{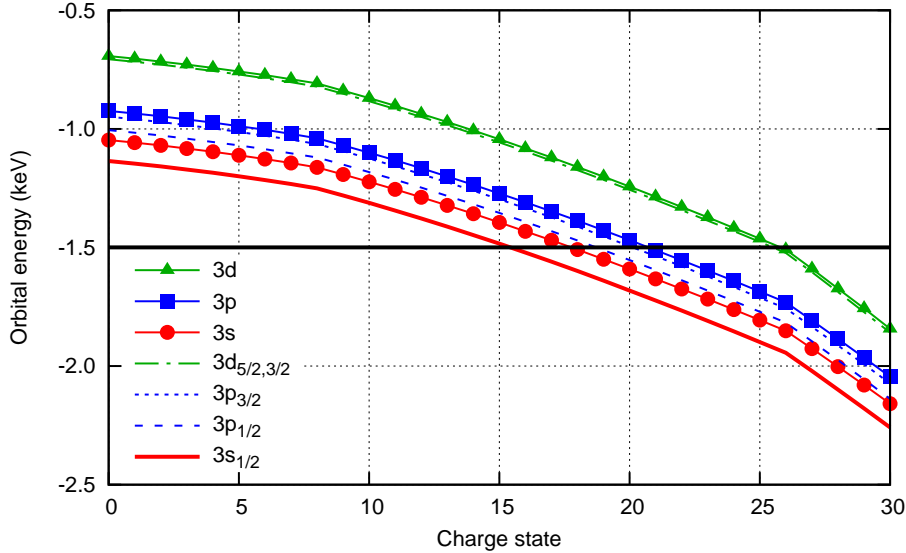} 
\caption{$M$-shell orbital energies of the ground configuration of Xe ions with and without relativistic effects.}
\label{fig:orbital_E@1.5keV}
\end{figure}

At a photon energy of 1.5~keV, $M$-shell electrons of neutral Xe can be ionized.
The $M$-shell photoionization stops as the charge increases when the $M$-shell ionization potential of the charge state is higher than the photon energy.
After this point, resonant photoexcitation can occur and Xe ions can further ionize via REXMI~\cite{RS2012,RR2013}.
Figure~\ref{fig:orbital_E@1.5keV} shows the $M$-shell orbital energies of the ground configuration of Xe ions as a function of charge state.
The non-relativistic results are plotted with lines and symbols, while the relativistic results are plotted with lines only.
The relativistic $3s_{1/2}$ orbital energy is lower than the non-relativistic $3s$ orbital energy by about 100~eV.
The spin-orbit splitting for $3p$ is about 100~eV, but the relativistic effects on $3d$ are less than 20~eV up to Xe$^{30+}$.
For the $M$-shell orbital energies, there are no dramatic changes due to relativistic effects, in contrast to the $L$-shell case in the previous subsection.

\begin{figure}
\includegraphics[width=0.75\textwidth]{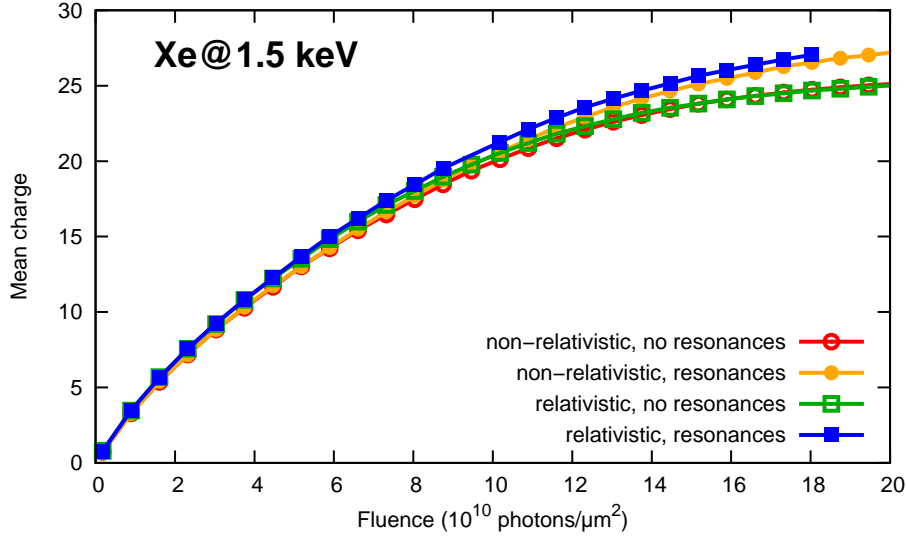} 
\caption{Mean charge of Xe at 1.5~keV as a function of fluence.}
\label{fig:mean_charge@1.5keV}
\end{figure}

Figure~\ref{fig:mean_charge@1.5keV} depicts the mean charge of Xe at 1.5~keV as a function of fluence.
The pulse duration is 80~fs FWHM.
The x-ray energy bandwidth is 1\% (15~eV) for the resonance calculations.
All computational parameters used are listed in Table~\ref{tab:comp_param}.
The fluence spans up to $2\times10^{11}$~photons/$\mu$m$^2$, which is higher than the one-photon absorption saturation fluence of neutral Xe at 1.5~keV ($\sim 1.1\times10^{10}$~photons/$\mu$m$^2$).
In the low fluence regime, the relativistic calculations (green and blue squares) are a bit higher than the non-relativistic calculations (red and orange circles).
But the mean charges with resonances (orange filled circles and blue filled squares) exceed those without resonances in the high fluence regime, due to REXMI~\cite{RS2012,RR2013}.
At this photon energy with higher fluences, it is expected that the resonance effect is more pronounced than the relativistic effect.

\begin{figure}[t]
\includegraphics[width=0.75\textwidth]{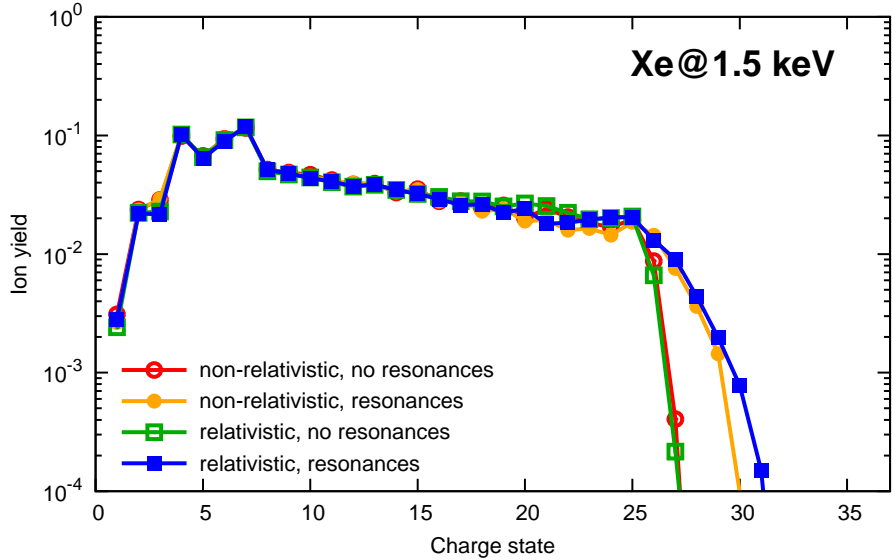} 
\caption{Charge-state distributions of Xe at 1.5~keV, after volume integration with a peak fluence of $1.3\times 10^{11}$ photons/$\mu$m$^2$.}
\label{fig:CSD@1.5keV}
\end{figure}

The calculated CSDs of Xe at 1.5~keV are shown in Fig.~\ref{fig:CSD@1.5keV}.
A peak fluence of $1.3\times10^{11}$~photons/$\mu$m$^2$ is used for volume integration, assuming a Gaussian beam profile.
Let us compare the non-relativistic cases with and without resonances (red open circles vs.\ orange filled circles).
Without resonances, the predicted highest charge state is $+27$, because the direct one-photon ionization from $M$-shell is closed at Xe$^{26+}$.
Note that production of charge states higher by $+1$ or $+2$ is possible via multiple-core-hole states, but their yields are quite low.
Including the REXMI mechanism dramatically enhances the yields of high charge states, similar to Ref.~\cite{HB2014}.
Next, we consider the relativistic effect without resonance (red open circles vs.\ green open squares).
As expected, there is no significant change in the relativistic calculation compared to the non-relativistic calculation (see Fig.~\ref{fig:orbital_E@1.5keV}).

However, our calculation with both relativistic treatment and resonant excitation illustrates that there is an interplay between the two effects.
The relativistic calculation with resonances (blue filled squares) clearly shows higher populations than the non-relativistic calculation with resonances (orange filled circles) for highly charged ions beyond the direct one-photon ionization limit (+26).
In the relativistic case, the REXMI mechanism starts earlier and the spin-orbit splittings increase chances to hit resonances.
Therefore, relativistic effects can further enhance the REXMI effect, producing more high charge states.

\section{Conclusion and perspective}\label{sec:conclusion}

\begin{figure}[t]
\includegraphics[width=0.75\textwidth]{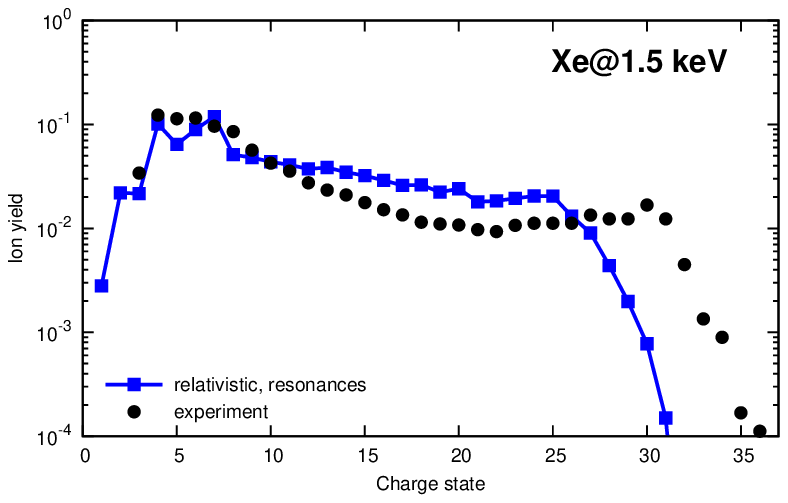} 
\includegraphics[width=0.75\textwidth]{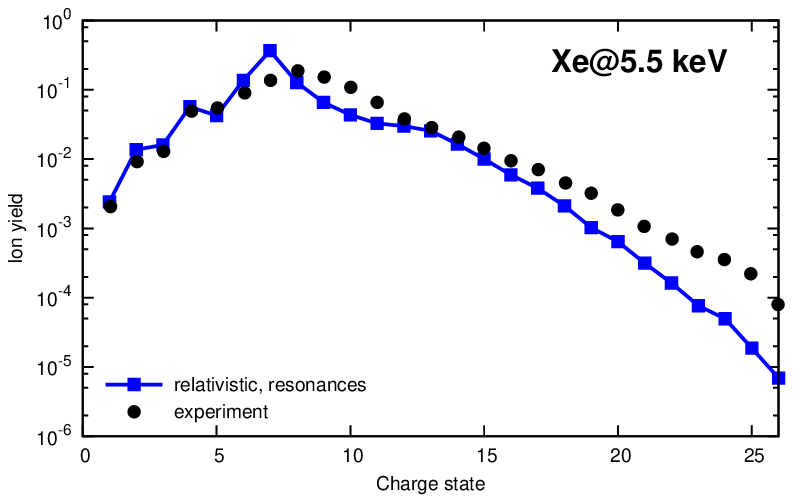} 
\caption{Charge-state distribution of Xe at 1.5~keV and 5.5~keV compared with experimental data~\cite{RS2012,FS2013}.}
\label{fig:CSD_compare}
\end{figure}

In this paper, we have extended the \textsc{xatom} toolkit to a relativistic version in order to further study x-ray multiphoton ionization dynamics of Xe atoms in XFEL beams. 
This extension is considered to be important because relativistic effects on ion CSDs could become significant in heavy atoms.
Our approach is to introduce leading-order relativistic energy corrections via perturbation theory, and cross sections and rates are correspondingly reformulated. 
We have confirmed that the calculated energy corrections, cross sections, and rates are in good agreement with the literature.
Also we have extended \textsc{xatom} to include resonant photoexcitation processes in both non-relativistic and relativistic treatments.

We have calculated CSDs of Xe atoms after interacting with intense x-ray pulses at $5.5$~keV and $1.5$~keV, respectively.
At the former photon energy, the ionization dynamics are influenced by relativistic effects, because deep inner-shell electrons are initially ionized.
On the other hand, at the latter photon energy, the resonance effect plays a particularly important role in the ionization dynamics at high x-ray intensity.
By using the extended \textsc{xatom} toolkit, we have examined the relativistic effect and the resonance effect separately, and have found a synergy effect when both of them are applied together in our calculations.
We have demonstrated that, generally speaking, both effects must be taken into account in x-ray multiphoton ionization dynamics calculations.
But do these effects resolve all discrepancies with experiment?

In Fig.~\ref{fig:CSD_compare}, we compare the experimental and present theoretical Xe CSDs, at 1.5~keV~\cite{RS2012} and 5.5~keV~\cite{FS2013}, respectively. 
The theory results shown include both relativistic and resonance effects. 
For this comparison we have performed the volume integration based on the x-ray beam parameters 
that were determined in each experiment.
Our theoretical prediction still underestimates the yields of highly charged ions in the CSD at both photon energies.
This might be due to our mean-field approach and a lack of higher-order many-body processes such as shakeoff processes and/or double Auger decays in our theoretical approach.
The discrepancy between theory and experiment might also be attributed to uncertainties in the x-ray beam parameters, such as the spatial beam profile and spectral bandwidth.
Therefore, for a more quantitative understanding of x-ray multiphoton ionization dynamics, it is desirable to improve the theoretical treatment for x-ray-induced processes and the calibration of the x-ray beam parameters in experiment.

\section{Acknowledgment}\label{sec:acknowledgment}
We thank Phay Ho for helpful discussions.
In the early stages of this project, Ekaterina Kuzmina participated as a summer student.


\end{document}